\documentclass[aps,preprint]{revtex4}
\begin{document}
\title{Decoherence due to three-body loss and its effect on the state of a Bose-Einstein condensate} 
\author{Michael W. Jack}
\surname{Jack}

\affiliation{NTT Basic Research Laboratories, NTT Corporation, 3-1, Morinosato Wakamiya, Atsugi-shi, Kanagawa 243-0198, Japan}

\begin{abstract}
A Born-Markov master equation  is used to investigate the decoherence of the state of a macroscopically occupied mode of a cold atom trap due to  three-body loss. In the large number limit only coherent states remain pure for times longer than the decoherence time: the time it takes for just three atoms to be lost from the trap.  For large numbers of atoms ($N>10^4$) the decoherence time  is found to be much faster than the phase collapse time caused by intra-trap atomic collisions. 
\end{abstract}
\maketitle 
Decoherence due to environmental coupling \cite{zurek91,caldeira83,walls85} can be very  rapid and has the effect of making certain pure quantum states unobtainable for all practical purposes. This means that the environment largely determines the state of the system as only so called ``robust states''  survive for long periods of time.
This is also true for cold trapped atoms and in particular when a Bose-Einstein condensate is present in the trap \cite{anderson95}. Even at zero temperature the condensate atoms are coupled to the free modes outside the trap by atomic loss processes and so these free modes act as an environment for the condensate. It has been argued \cite{barnett96,shimizu00} that environmental coupling will leave the condensate in a coherent state with a well-defined phase due to the robustness of the coherent states.

The form of the environmental coupling is crucial in determining the robust states of an open system \cite{zurek91}. It is well known that the coherent states are robust under  linear coupling \cite{walls85,zurek93},
or single-body loss, in the sense that an initial coherent state decays in amplitude but otherwise remains unchanged. More important for large Bose-Einstein condensates is loss due to three-body recombination \cite{kagan85,burt97, soding99,esry99}, which limits the lifetime and size of the present condensates. Three-body loss can be considered to be a fundamental loss process for
trapped atoms as it cannot be reduced (except by changing the scattering length \cite{inouye98}), unlike loss due to spin-relaxation which does not occur in an optical trap and unlike loss due to collisions with untrapped atoms which can be reduced by improving the vacuum.  In this letter I aim to show that three-body loss, gives rise to a very rapid decoherence of certain states in spite of the fact that it leads to only a slow decay in the number of  trapped atoms. In the large number limit  the robust states are similar to coherent states with a well-defined phase. 

In opposition to the above argument, elastic atom-atom collisions within the condensate will give rise to a dephasing between states of different number and in the absence of environmental coupling lead to a phase-collapse \cite{wright96,lewenstein96,villain97}. However, I will show that the timescale at which the  three-body loss decoheres an initial state to the robust states is much more rapid than the phase collapse for large condensates. The collisions are then more likely to lead to a slow diffusion of a well-defined phase rather than a collapse.

To investigate the quantum properties of three-body loss 
I derive a Born-Markov master
equation for the density matrix of the trapped atoms.    Born-Markov master equations have been an important tool for describing
radiative decay in quantum optics (see for example \cite{dansbook}).

For cold atoms in a
trap the three-body recombination process produces a molecule and a single atom,  (e.g. Rb $+$ Rb $+$ Rb $\rightarrow$ Rb $+$ Rb$_{2}$), where the  molecule is an atom-atom bound state. The weakest binding energy, $E_{b}$, is related to the $s$-wave
scattering length, $a$, by $E_{b}=\hbar\omega_{b}=\hbar^{2}/m a^2$ \cite{sakurai}. In this work I consider only the repulsive interaction case, $a>0$.
Due to the large binding energy and momentum conservation the molecule
and atom are both created with a large kinetic energy and escape from the trap. We are then interested in the influence of this loss
process on the behavior of the remaining trapped atoms.

For low-temperature atoms the three-body process becomes a point
interaction characterized by a single parameter $\kappa$ such that the
Hamiltonian can be written
\begin{equation}
  \label{eq:interactionHamiltonian}
  H_{I}=\hbar\kappa\int d{\mathbf x} \left[\hat{\psi}^{3}_{\rm T}({\mathbf x})\hat{\psi}^{\dagger}_{\rm F}({\mathbf x})\hat{\phi}_{\rm F}^{\dagger}({\mathbf x})+{\rm h.c.}\right],
\end{equation}
where $\hat{\psi}_{\rm T}({\mathbf x})$, $\hat{\psi}_{\rm F}({\mathbf x})$ and
$\hat{\phi}_{\rm F}({\mathbf x})$ are the field operators for the trapped atoms,
the free (untrapped) atoms and free molecules, respectively. The field operators
commute with each other but have the usual Bose commutation relations
with their Hermitian conjugate.

The Hamiltonian for the cold trapped atoms, $H_{\rm T}$, is the usual many-body Hamiltonian describing two-body $s$-wave  scattering  (see, for example, the review article \cite{dalfovo99}). In most circumstances three-body
{\em elastic} collisions will be negligible in comparison to two-body
elastic collisions so I do not include them here.

The density matrix of the total system satisfies the interaction-picture equation of motion
$d\rho_{\rm tot}(t)/dt=-i[H_{I}(t),\rho_{\rm tot}(t)]/\hbar$,
  where $H_{I}(t)=e^{-i(H_{\rm T}+H_{\rm F})t/\hbar}H_{I}e^{i(H_{\rm T}+H_{\rm F})t/\hbar}$ and  $H_{\rm F}$ is the Hamiltonian of the free particles. The free particles and the trapped atoms are assumed to be initially
  uncoupled with the states outside the trap unpopulated;
  $\rho_{\rm tot}(0)=\nu_{1}\otimes\nu_{2}\otimes\rho(0)$, where $\rho(0)$,
  $\nu_{2}=|\{0\}\rangle\langle\{0\}|_{\phi_{\rm F}}$ and
  $\nu_{1}=|\{0\}\rangle\langle\{0\}|_{\psi_{\rm F}}$ are the density matrices of
  the trapped atoms, free molecules and free atoms, respectively.

  Assuming a low loss rate we can expand to second order in $\kappa$
  (see, for example, \cite{dansbook})  and the equation of motion
  for the reduced density matrix  $\rho={\rm tr}_{\rm F}\left\{\rho_{\rm tot}\right\}$, becomes 
\begin{equation}
  \label{eq:reducedDensityMatrix2}
\frac{d\rho}{dt}=-\!\!\int^{t}_{0}\!\!\!ds\!\!\int \!\!d{\mathbf x}d{\mathbf y}\; f({\mathbf x},{\mathbf y};t,s)[\hat{\psi}^{\dagger 3}_{\rm T}({\mathbf x}t),\hat{\psi}^{3}_{\rm T}({\mathbf y}s)\rho]+\mbox{h.c}.
\end{equation}
where
\begin{equation} 
f({\mathbf x},{\mathbf y};t,s)=\kappa^{2}\langle\{0\}|\hat{\phi}_{\rm F}\!({\mathbf x}t)\hat{\psi}_{\rm F}\!({\mathbf x}t)\hat{\psi}^{\dagger}_{\rm F}\!({\mathbf y}s)\hat{\phi}_{\rm F}^{\dagger}\!({\mathbf y}s)|\{0\}\rangle.
\end{equation}
Note that the first term in the perturbation, ${\rm tr}_{\rm F}\left\{\left[H_{I}(s),\rho_{\rm tot}(t)\right]\right\}=0$.
Following Kagan {\em et al.}\cite{kagan85}, we can expand the field operators for the free particles as plane waves modes with the energies $\hbar^{2} k^{2}_{1}/2m$ and $\hbar^{2} k_{2}^{2}/4m-E_{b}$ for the atom and the molecule, respectively (due to the high velocity of the free particles  we can neglect the effects of gravity). Since the shortest length scale of the trapped atoms is given by the de
Broglie wavelength, on the length scale of the emitted particles ($\sim a$), there is no
variation of the field of the trapped atoms and we can make the
replacement $\hat{\psi}_{\rm T}({\mathbf y})\rightarrow\hat{\psi}_{\rm T}({\mathbf
  x})$ in Eq.(\ref{eq:reducedDensityMatrix2}). 
 We can then integrate
the memory function over ${\mathbf y}$  which gives rise to the delta function $\delta({\mathbf k}_{1}+{\mathbf k}_{2})$, expressing the equal an opposite momentum of the molecule and atom. We then perform the integration over ${\mathbf k}_{2}$ to give,
\begin{equation}
 F(t-s) =\frac{2\kappa^{2}}{(2\pi)^2}\int k_{1}^2 dk_{1}\: e^{-i\left(\frac{3\hbar k_{1}^2}{4m}-\omega_{b}\right)(t-s)}\label{integral},
\end{equation}
where $F(t-s)=\int d{\bf y}f({\mathbf x},{\mathbf y};t,s)$. Note that the ${\mathbf x}$ dependence has disappeared. Replacing the variable of integration by
$K=3\hbar k_{1}^2/4m$ in Eq.(\ref{integral}) and noting that we are only interested in
frequencies in the range $\omega_{b}-\Omega$ to $\omega_{b}+\Omega$,
where $E_{b}/\hbar\gg\Omega\gg\kappa$, the integral can be approximated by
\begin{equation}
 F(t-s)\approx \frac{\gamma}{6\pi}\int ^{\omega_{b}+\Omega}_{\omega_{b}-\Omega}dK\: e^{-i\left(K-\omega_{b}\right)(t-s)},
\end{equation}
where $\gamma=3(\kappa^{2}/2\pi)(4m/3\hbar)^{3/2}\sqrt{\omega_{b}}$. As $ \Omega$ is much larger than
any system frequencies we can  write
$F(t-s)\approx\gamma\delta(t-s)/3$ and the final Born-Markov master equation for the reduced density matrix
of the trapped atoms is given by
\begin{widetext}
\begin{equation}
\label{eq:masterEquation}
 \frac{d\rho}{dt}=-\frac{i}{\hbar}[H_{\rm T},\rho(t)]+\frac{\gamma}{6} \int d{\mathbf x}\left[2\hat{\psi}^{3}({\mathbf x})\rho\hat{\psi}^{\dagger 3}({\mathbf x})-\hat{\psi}^{\dagger 3}({\mathbf x})\hat{\psi}^{3}({\mathbf x})\rho-\rho\hat{\psi}^{\dagger 3}({\mathbf x})\hat{\psi}^{3}({\mathbf x})\right],
\end{equation}
\end{widetext}
where I have dropped the
subscript on the field operator and returned to the Schr\"{o}dinger
picture.   This equation is a course-grained equation, valid for  time scales of the
trapped atoms
$\gg 2\pi/\omega_{b}$ and length scales  $\gg a$. By calculating the number decay via Eq.(\ref{eq:masterEquation}) we can determine that $\gamma$ is related to the recombination event rate $K_{3}$ of Ref.\cite{esry99} by $K_{3}=12\gamma$.

This master equation can be used to derive the result of Kagan {\em et al.} \cite{kagan85} but it is
much more general as it can be used to calculate the behavior of any
expectation value of the trap operators. In this
work, however, I am interested in the effect of the three-body decay
on the quantum state of the condensate. 

For a weakly-interacting gas at zero temperature, the atoms may be assumed to be mostly in the condensate. A condensate of $n$ atoms has a wave function that satisfies the time-independent  Gross-Pitaevskii equation. In the Thomas-Fermi approximation (see \cite{dalfovo99} and references within) the solution is: $gn|\psi_{n}({\mathbf x})|^2 \approx \mu_{n}-V_{\rm trap}({\mathbf x})$,
for all ${\mathbf x}$ where  $V_{\rm trap}({\mathbf x})<\mu_{n}$ and zero elsewhere. The chemical potential $\mu_{n}$ is determined from the normalization of the wavefunction $\psi_{n}$. At higher temperatures the noncondensate particles occupying higher energy modes are expected to also act as an environment for the condensate mode. The lack of a suitable theory  to describe the quantum effects of this interaction (see Ref.\cite{gardiner97} for an attempt)  is particularly acute in this case as this interaction is expected to compete with the loss from the trap in decohering the state of the condensate.

Let us assume that the initial condensate states of interest are sharply peaked about the mean number $N$ such that $|N-n|/N \ll 1$ and during the short time period of interest (the time it takes to lose
$\sim 3$ atoms) the system stays within this linear region.  Due to the fact that the wavefunction is stable under small changes in number for $a>0$ we can make the approximations  
$\psi_{n}({\mathbf x})\approx \psi_{N}({\mathbf x})$ and $\mu_{n}\approx\mu_{N}$,
see Ref.\cite{wright96}. The $n$ dependence of $\mu_{n}$ will be discussed latter.  Under these approximations we
can make the replacement $\hat{\psi}({\mathbf x})\rightarrow\psi_{N}({\mathbf
  x})\hat{A}$ in Eq.(\ref{eq:masterEquation}), where $\hat{A}$ is
the annihilation operator for the condensate mode. In order to describe longer times where a significant portion of atoms are lost from the condensate a more careful approach to defining the condensate mode may be necessary, (see Ref.\cite{shimizu00}). In general, the changing number will give rise to
excitations but we can neglect these effects for slow decaying numbers
over short times.  

With the above assumptions, the master equation for the condensate mode becomes
\begin{equation}
\label{eq:masterEquation_singleMode}
 \frac{d\rho}{dt}=\frac{\tilde{\gamma}}{6}\left[2\hat{A}^{3}\rho\hat{A}^{\dagger 3}-\hat{A}^{\dagger 3}\hat{A}^{3}\rho-\rho\hat{A}^{\dagger 3}\hat{A}^{3}\right],
\end{equation}
where $\tilde{\gamma}=\gamma \int d{\mathbf x}|\psi_{N}({\mathbf x})|^6$. In Eq.(\ref{eq:masterEquation_singleMode}) I have assumed that the free Hamiltonian has the form $H_{\rm T}=\hbar\mu_{N} \hat{A}^{\dagger}\hat{A}$ and that we are working in the a frame rotating at the frequency defined by $\mu_{N}$.
This equation represents a nonlinear damping of the condensate mode and in the  large number limit it has an analytical solution. 

Due to the fact that the operator $\hat{A}^{3}$ couples only every third number state,  there are $3\times 3$ independent sub-manifolds of the density matrix corresponding to the number states $|m;j\rangle\equiv|3m+j\rangle$, where $m$ is an integer and $j=0,1$ or $2$ are the labels for the sub-manifolds. 
 The master equation yields an equation  for each of the $3\times 3$ the density matrices $\rho_{jk}$ $(j,k=0,1$ or $2)$ of the sub-manifolds of the form
\begin{equation}
\label{eq:masterEqn_singleMode_subManifold}
 \frac{d\rho_{jk}}{dt}\approx\Gamma\left[2\hat{a}_{j}\rho_{jk}\hat{a}^{\dagger}_{k}-\hat{a}^{\dagger}_{j}\hat{a}_{j}\rho_{jk}-\rho_{jk}\hat{a}^{\dagger}_{k}\hat{a}_{k}\right]
\end{equation}
 where $\Gamma =3N^2\tilde{\gamma}/2$. Assuming that the number distribution is sharply peaked at a large number the  mode operators for each of the three manifolds can be written as 
\begin{equation}
\hat{a}_{j}=\sum_{\rm lin.}\sqrt{C_{N}^{j}+m+1}|m;j\rangle\langle m+1;j|+O(N^{0}),
\end{equation}
where $C_{N}^{j}=(-2N+3j-3)/9$, and the sum is over states in the linear region about $N$. The  mode operators satisfy $[\hat{a}_{j},\hat{a}^{\dagger}_{j'}]=  \delta_{j,j'}$,
where I have assumed that only number states in the linear region about $N$ are occupied and so it is possible to shift the number states up by one. The equations (\ref{eq:masterEqn_singleMode_subManifold}) are linear master equations for which their exists a well known solution in terms of the coherent states  \cite{walls85}.

We can define coherent states for the mode operators $\hat{a}_{j}$  in the linear region about $N$ as
\begin{equation}
|\alpha_{j};j\rangle=\mathcal{N}\sum_{\rm lin.}\frac{\alpha_{j}^{C_{N}^{j}+m}}{\sqrt{(C_{N}^{j}+m)!}}|m;j\rangle, \label{coherentState}
\end{equation}
where $\alpha_{j}$  are complex numbers. These states satisfy $\hat{a}_{j} |\alpha_{j};j\rangle\approx\alpha_{j}|\alpha_{j};j\rangle$. They are a superposition of every third number state and the  number distribution is given by $P(n)=|\langle n|\alpha_{j};j\rangle|^2=\delta_{n,3m+j}Q(m)$, where, in the large number limit, we can show that
\begin{eqnarray}
Q(m)&\approx &\frac{\mathcal{N}^{2}e^{N/9}}{\sqrt{2\pi (N/9)}}\left[\frac{9|\alpha_{j}|^2}{N}\right]^{C_{N}^{j}+m}e^{-\frac{(3m+j-N)^2}{2N}},\label{numberDistribution}
\end{eqnarray}
using the relation $x! x_{0}^{-x}\approx \sqrt{2\pi x_{0}}\:e^{\frac{(x-x_{0})^2}{2x_{0}}-x_{0}}$
 for $x_{0}\gg 1$ and $|x-x_{0}|/x_{0}\ll 1$  \cite{integrals-series-products}. The states that we are interested in are peaked at $n=N$ so  $|\alpha_{j}|^2= N/9$ and from Eq.(\ref{numberDistribution}) we can see that the normalization factor in the large number limit is $\mathcal{N}\approx e^{-|\alpha_{j}|^{2}/2}$.

The evolution of an arbitrary element of the density matrix in this
coherent state representation, $|\alpha;j\rangle\langle\beta;k|$,  is \cite{walls85}:
\begin{equation}
(|\alpha;j\rangle\langle\beta;k|)_{t}\approx \langle\alpha;j|\beta;j\rangle^{1-e^{-2\Gamma t}}\left|\alpha e^{-\Gamma t};j\right\rangle\left\langle\beta e^{-\Gamma t};k\right|,\label{solution1}
\end{equation}
where $\langle\alpha;j|\beta;j\rangle\approx\exp\left[-(|\alpha|^2+|\beta|^{2}-2\alpha\beta^{*})/2\right]$. For $j=k$ this equation shows that any off-diagonal elements in the coherent-state basis, $|\alpha;j\rangle$, rapidly dephase. This means that an
initial pure state which is a superposition of the states $|\alpha;j\rangle$ will rapidly become a mixed state. For example, the superposition of two states with different phases: \begin{equation}
(|\alpha\rangle+|\alpha e^{i\phi}\rangle)(\langle\alpha|+\langle\alpha e^{i\phi}|)\longrightarrow |\alpha\rangle\langle \alpha|+|\alpha e^{i\phi}\rangle\langle\alpha e^{i\phi}|
\end{equation}
at the rate $\sim |\alpha|^2\Gamma$. This can be seen by noting that $\langle\alpha|\alpha e^{i\phi}\rangle=\exp\{-2i|\alpha|^2 e^{i\phi/2}\sin(\phi/2)\}$ and therefore the rate at which the off-diagonal terms dephase is $\sim |\alpha|^2\Gamma$.  A number state, which is also a superposition of coherent states of different phase, will quickly evolve into a mixed state with a binomial number distribution over the number states.
For $k\neq j$, Eq.(\ref{solution1}) shows that any state where the coherent amplitude $\alpha$ is not the same in all the  sub-manifolds will also rapidly dephase.

The robust states then have the general form $|R(\alpha)\rangle=\sum_{j} c_{j}|\alpha;j\rangle$,
where $\sum_{j}|c_{j}|^2=1$.  As the preparation of the initial condensate is via processes that are not likely to preserve the independence of the submanifolds, i.e. evaporative cooling, the most relevant states are those where $c_{j}=1/\sqrt{3}$. In this case,  the robust states  have the form 
\begin{eqnarray}
|R(\alpha)\rangle
&=&\mathcal{N}\sum_{{\rm lin.}}\frac{\alpha^{C^{0}_{N}+n/3}}{\sqrt{3\Gamma (C^{0}_{N}+n/3+1)}}|n\rangle,
\end{eqnarray}
where $\Gamma(x)$ is the gamma function. These states have the property $\hat{A}^{3}|R(\alpha)\rangle=3N\alpha|R(\alpha)\rangle$ in the linear regime. In general, $|R(\alpha)\rangle$ is not an eigenstate of $\hat{A}$ but in the large-number limit,  it is possible to show that $\langle \beta|R(\alpha)\rangle\approx e^{iC_{N}^{0}\theta}\langle \beta|(3N^2\alpha)^{1/3}\rangle$,
where $ \alpha=|\alpha|e^{i\theta}$ and $|\beta\rangle$ is an ordinary coherent state. And so, in the large-number limit, the robust states become equivalent to coherent states  (except for a phase factor) with the amplitude $(3N^2\alpha)^{1/3}$.

From Eq.(\ref{solution1}) and $|\alpha|^2\approx N/9$, the time scale of the decoherence is
\begin{equation}
\tau_{\rm decoh}^{-1}\sim \tilde{\gamma}N^3/3,
\end{equation}
where $\tilde{\gamma}=[(15\lambda)^{4/5}/14^3\pi^2]K_{3}(a_{\perp}^{4} a N)^{-6/5}$. In determining $\tilde{\gamma}$ I have used the Thomas-Fermi solution of the Gross-Pitaevskii  equation for the cylindrically-symmetric harmonic trap: $V_{\rm T}({\mathbf x})=\frac{1}{2}m \omega_{\perp}^{2}(r^2+\lambda^{2} z)$, where $m$ is the atomic mass and $\omega_{\perp}=\hbar/m a^{2}_{\perp}$ is the radial trap frequency.
The decoherence rate is of the order of $N$ times the decay rate for the atom number, or, the time it takes to lose $\sim 3$ atoms from the trap.
For example, a $^{87}$Rb condensate of $3\times 10^5$ atoms with $K_{3}=2.2\times 10^{-28}$ cm$^6$/s and the trap parameters $\omega_{\perp}=2\pi \times 157$ Hz and $\lambda=0.075$ corresponding to the experiment described in \cite{soding99}, yields $\tilde{\gamma}\approx 3.3\times 10^{-12}$s$^{-1}$ and $\tau_{\rm decoh}\approx 4\times 10^{-5}$ s. 

This decoherence process could be measured directly by observing the rate of decay of visibility  of the interference of a superposition of states of different phases as a function of the lifetime of the superposition. Assuming the condensate is initially in a coherent state, a process similar to that described in Ref.\cite{hall98} could be used to create a superposition:  coherently transfer half the atoms to another spin state by a $\frac{\pi}{2}$ laser pulse; let the phases of the two spin states evolve independently and then transfer the atoms back to the original spin state with another $\frac{\pi}{2}$  pulse. 

To make any conclusions as to the state of the condensate, the above decoherence time scale must be compared with that of the collapse of the phase due to condensate-atom collisions calculated in   Ref.\cite{wright96}. Due to atom-atom collisions the chemical potential, $\mu_{n}$, is number dependent (this has been neglected in the present work so far) and leads to a dephasing of any superposition of number states. Assuming the same trap symmetry and the Thomas-Fermi limit of the Gross-Pitaevskii equation as above,  the collapse time for a number variance of $\sqrt{N}$ is given by \cite{wright96} $\tau_{\rm coll}\approx (5/\omega_{\perp})\left(a_{\perp}/\lambda a\right)^{2/5}N^{1/10}$. The same experimental parameters considered above give a collapse time of $ \tau_{\rm coll}\approx 0.4$s, or $4$ orders of magnitude longer than the decoherence time. In general, $\tau_{\rm coll}/\tau_{\rm decoh}\propto O(N^{21/10})$ and a rough calculation shows that the decoherence process will dominate for $> 10^4$ atoms. A more detailed analysis is necessary but this suggests that collisions will lead a slow phase diffusion of a well-defined phase rather than a phase collapse. 

In this work I have presented a realistic mechanism for decoherence at low temperatures which supports the  assumption that the state of the condensate is a coherent state with a well-defined phase.  One can also understand the experimental finding that the condensate phase is rather robust\cite{hall98}; the environment, rather than diffusing the phase helps to keep the phase well-defined. However, this work also suggests directions for investigation that go beyond the assumption of a well-defined phase. In particular,  the observation of macroscopic-superpositions of different phase and their decay.  In contrast to condensed matter systems where there is often a proliferation of possible  decoherence mechanisms that cannot be isolated from each other, atomic Bose-Einstein condensates are relatively clean systems with few possible decoherence mechanisms and are thus a promising test bed for understanding decoherence and quantum superpositions in general.  Atom loss (single body loss) has been shown to have an adverse effect on quantum entanglement between atoms\cite{sorensen}. Therefore trap loss (in particular, three-body loss) has important consequences for  the possible role of Bose-Einstein condensates in quantum information technology.


I would like to thank  Makoto Yamashita for many useful discussions. 

\bibliographystyle{apsrev}

\begin{thebibliography}{27}
\expandafter\ifx\csname natexlab\endcsname\relax\def\natexlab#1{#1}\fi
\expandafter\ifx\csname bibnamefont\endcsname\relax
  \def\bibnamefont#1{#1}\fi
\expandafter\ifx\csname bibfnamefont\endcsname\relax
  \def\bibfnamefont#1{#1}\fi
\expandafter\ifx\csname citenamefont\endcsname\relax
  \def\citenamefont#1{#1}\fi
\expandafter\ifx\csname url\endcsname\relax
  \def\url#1{\texttt{#1}}\fi
\expandafter\ifx\csname urlprefix\endcsname\relax\def\urlprefix{URL }\fi
\providecommand{\bibinfo}[2]{#2}
\providecommand{\eprint}[2][]{\url{#2}}

\bibitem[{\citenamefont{Zurek}(1991)}]{zurek91}
\bibinfo{author}{\bibfnamefont{W.~H.}~\bibnamefont{Zurek}},
  \bibinfo{journal}{Phys. Today} \textbf{\bibinfo{volume}{44}},
  \bibinfo{pages}{36} (\bibinfo{year}{1991});
\bibinfo{author}{\bibfnamefont{W.~H.}~\bibnamefont{Zurek}},
  \bibinfo{journal}{Phys. Rev. D} \textbf{\bibinfo{volume}{24}},
  \bibinfo{pages}{1516} (\bibinfo{year}{1981});\bibinfo{author}{\bibfnamefont{W.~H.}~\bibnamefont{Zurek}},
  \bibinfo{journal}{Phys. Rev. D} \textbf{\bibinfo{volume}{26}},
  \bibinfo{pages}{1862} (\bibinfo{year}{1982}).

\bibitem[{\citenamefont{Caldeira}(1983)}]{caldeira83}
\bibinfo{author}{\bibfnamefont{A.~O.}~\bibnamefont{Caldeira}} \bibnamefont{and}
\bibinfo{author}{\bibfnamefont{A.~J.}~\bibnamefont{Leggett}},
  \bibinfo{journal}{Physica (Utrecht)} \textbf{\bibinfo{volume}{121A}},
  \bibinfo{pages}{587} (\bibinfo{year}{1983});
\bibinfo{author}{\bibfnamefont{A.~O.}~\bibnamefont{Caldeira}} \bibnamefont{and}
\bibinfo{author}{\bibfnamefont{A.~J.}~\bibnamefont{Leggett}},
  \bibinfo{journal}{Phys. Rev. A} \textbf{\bibinfo{volume}{31}},
  \bibinfo{pages}{1059} (\bibinfo{year}{1985}).
\bibitem[{\citenamefont{Walls and Milburn}(1985)}]{walls85}
\bibinfo{author}{\bibfnamefont{D.~F.}~\bibnamefont{Walls}} \bibnamefont{and}
  \bibinfo{author}{\bibfnamefont{G.~J.}~\bibnamefont{Milburn}},
  \bibinfo{journal}{Phys. Rev. A} \textbf{\bibinfo{volume}{31}},
  \bibinfo{pages}{2403} (\bibinfo{year}{1985}).

\bibitem[{\citenamefont{Anderson et~al.}(1995)\citenamefont{Anderson, Ensher,
  Matthews, Wieman, and Cornell}}]{anderson95}
\bibinfo{author}{\bibfnamefont{M.~H.}~\bibnamefont{Anderson}},
  {\em et al.}, \bibinfo{journal}{Science}
  \textbf{\bibinfo{volume}{269}}, \bibinfo{pages}{198} (\bibinfo{year}{1995});\bibinfo{author}{\bibfnamefont{C.~C.}~\bibnamefont{Bradley}},
  {\em et al.},
  \bibinfo{journal}{Phys. Rev. Lett.} \textbf{\bibinfo{volume}{75}},
  \bibinfo{pages}{1687} (\bibinfo{year}{1995});\bibinfo{author}{\bibfnamefont{K.~B.}~\bibnamefont{Davis}},
  {\em et al. },
  \bibinfo{journal}{Phys. Rev. Lett.} \textbf{\bibinfo{volume}{75}},
  \bibinfo{pages}{3969} (\bibinfo{year}{1995}).

\bibitem{barnett96}
\bibinfo{author}{\bibnamefont{S.~M.}~\bibnamefont{Barnett}} \bibnamefont{et al.},
\bibinfo{journal}{J. Res. Natl. Stand. Technol.} \textbf{\bibinfo{volume}{101}},
\bibinfo{pages}{593} (\bibinfo{year}{1996}).

\bibitem[{\citenamefont{Shimizu and Miyadera}(2000)}]{shimizu00}
\bibinfo{author}{\bibfnamefont{A.}~\bibnamefont{Shimizu}} \bibnamefont{and}
  \bibinfo{author}{\bibfnamefont{T.}~\bibnamefont{Miyadera}},
  \bibinfo{journal}{Phys. Rev. Lett.} \textbf{\bibinfo{volume}{85}},
  \bibinfo{pages}{688} (\bibinfo{year}{2000}).

\bibitem[{\citenamefont{Zurek et~al.}(1993)\citenamefont{Zurek, Habib, and
  Paz}}]{zurek93}\bibinfo{author}{\bibfnamefont{W.~H.}~\bibnamefont{Zurek}},
 {\em et al.}, \bibinfo{journal}{Phys. Rev. Lett.} \textbf{\bibinfo{volume}{70}},
  \bibinfo{pages}{1187} (\bibinfo{year}{1993});
\bibinfo{author}{\bibfnamefont{M.~R.}~\bibnamefont{Gallis}},\bibinfo{journal}{Phys. Rev. A} \textbf{\bibinfo{volume}{53}},
  \bibinfo{pages}{655} (\bibinfo{year}{1996}).




\bibitem[{\citenamefont{Kagan et~al.}(1985)\citenamefont{Kagan, Svistunov, and
  Shlyapnikov}}]{kagan85}
\bibinfo{author}{\bibfnamefont{Y.}~\bibnamefont{Kagan}}, {\em et al.},
  \bibinfo{journal}{JETP Lett.}
  \textbf{\bibinfo{volume}{42}}, \bibinfo{pages}{209} (\bibinfo{year}{1985}).

\bibitem[{\citenamefont{Burt et~al.}(1997)\citenamefont{Burt, Ghrist, Myatt,
  Holland, Cornell, and Wienman}}]{burt97}
\bibinfo{author}{\bibfnamefont{E.~A.}~\bibnamefont{Burt}},
  {\em et al.}, \bibinfo{journal}{Phys. Rev. Lett.}
  \textbf{\bibinfo{volume}{79}}, \bibinfo{pages}{337} (\bibinfo{year}{1997}).

\bibitem[{\citenamefont{S\"{o}ding et~al.}(1999)\citenamefont{S\"{o}ding,
  Gu\'{e}ry-Odelin, anf F.~Chevy, Inamori, and j.~Dalibard}}]{soding99}
\bibinfo{author}{\bibfnamefont{J.}~\bibnamefont{S\"{o}ding}}, {\em et al.}, \bibinfo{journal}{Appl. Phys. B}
  \textbf{\bibinfo{volume}{69}}, \bibinfo{pages}{257} (\bibinfo{year}{1999}).

\bibitem[{\citenamefont{Esry et~al.}(1999)\citenamefont{Esry, Greene, and James
  P.~Burke}}]{esry99}
\bibinfo{author}{\bibfnamefont{B.~D.}~\bibnamefont{Esry}}, {\em et al.},
   \bibinfo{journal}{Phys. Rev. Lett.}
  \textbf{\bibinfo{volume}{83}}, \bibinfo{pages}{1751} (\bibinfo{year}{1999}).

\bibitem[{\citenamefont{Inouye et~al.}(1998)\citenamefont{Inouye, Mathews,
  Stenger, Miesner, Stamper-Kurn, and Ketterle}}]{inouye98}
\bibinfo{author}{\bibfnamefont{S.}~\bibnamefont{Inouye}}, {\em et al.} ,
  \bibinfo{journal}{Nature (London)} \textbf{\bibinfo{volume}{392}},
  \bibinfo{pages}{151} (\bibinfo{year}{1998}).



\bibitem[{\citenamefont{Wright et~al.}(1996)\citenamefont{Wright, Walls, and
  Garrison}}]{wright96}
\bibinfo{author}{\bibfnamefont{E.~W.}~\bibnamefont{Wright}}, {\em et al.}, 
  \bibinfo{journal}{Phys. Rev. Lett.} \textbf{\bibinfo{volume}{77}},
  \bibinfo{pages}{2158} (\bibinfo{year}{1996});
\bibinfo{author}{\bibfnamefont{E.~M.}~\bibnamefont{Wright}}, {\em et al.},
  \bibinfo{journal}{Phys. Rev. A} \textbf{\bibinfo{volume}{56}},
  \bibinfo{pages}{591} (\bibinfo{year}{1997}).
\bibitem{lewenstein96}
\bibinfo{author}{\bibnamefont{M.}~\bibnamefont{Lewenstein}} \bibnamefont{and}
\bibinfo{author}{\bibnamefont{L.}~\bibnamefont{You}},
\bibinfo{journal}{Phys. Rev. Lett.} \textbf{\bibinfo{volume}{77}},
\bibinfo{pages}{3489} (\bibinfo{year}{1996}).


\bibitem{villain97}
\bibinfo{author}{\bibnamefont{P.}~\bibnamefont{Villain}} {\em et al.},
\bibinfo{journal}{J. Mod. Opt.} \textbf{\bibinfo{volume}{44}},
\bibinfo{pages}{1775} (\bibinfo{year}{1997}).
\bibitem[{\citenamefont{Walls and Milburn}(1994)}]{dansbook}
\bibinfo{author}{\bibfnamefont{D.~F.}~\bibnamefont{Walls}} \bibnamefont{and}
  \bibinfo{author}{\bibfnamefont{G.~J.}~\bibnamefont{Milburn}}, \emph{\bibinfo{title}{Quantum Optics}} 
(\bibinfo{publisher}{Springer-Verlag}, \bibinfo{address}{New York}, \bibinfo{year}{1994}).


\bibitem[{\citenamefont{Sakurai}(1994)}]{sakurai}
\bibinfo{author}{\bibfnamefont{J.~J.}~\bibnamefont{Sakurai}},
  \emph{\bibinfo{title}{Modern Quantum Mechanics, Revised Edition}}
  (\bibinfo{publisher}{{Addison-Wesley Publishing Company, Inc.}},
  \bibinfo{address}{Massachusetts}, \bibinfo{year}{1994}).

\bibitem[{\citenamefont{Dalfovo et~al.}(1999)\citenamefont{Dalfovo, Giorgini,
  Pitaevskii, and Stringari}}]{dalfovo99}
\bibinfo{author}{\bibfnamefont{F.}~\bibnamefont{Dalfovo}}, {\em et al.},
  \bibinfo{journal}{Rev. Mod. Phys.} \textbf{\bibinfo{volume}{71}},
  \bibinfo{pages}{463} (\bibinfo{year}{1999}).


\bibitem{gardiner97}
\bibinfo{author}{\bibnamefont{C.~W.}~\bibnamefont{Gardiner}} \bibnamefont{and}
\bibinfo{author}{\bibnamefont{P.}~\bibnamefont{Zoller}},
\bibinfo{journal}{Phys. Rev. A} \textbf{\bibinfo{volume}{55}},
\bibinfo{pages}{2902} (\bibinfo{year}{1997}).



\bibitem[{\citenamefont{Gradshteyn and
  Ryzhik}(1994)}]{integrals-series-products}
\bibinfo{author}{\bibfnamefont{I.~S.}~\bibnamefont{Gradshteyn}}
  \bibnamefont{and} \bibinfo{author}{\bibfnamefont{I.~M.}~\bibnamefont{Ryzhik}}, \emph{\bibinfo{title}{Table of Integrals, Series, and
  Products}} (\bibinfo{publisher}{Academic Press Limited},
  \bibinfo{address}{London}, \bibinfo{year}{1994}), \bibinfo{edition}{5th} ed.



\bibitem{hall98}
\bibinfo{author}{\bibnamefont{D.~S.}~\bibnamefont{Hall}} {\em et al.},
\bibinfo{journal}{Phys. Rev. Lett.} \textbf{\bibinfo{volume}{81}},
\bibinfo{pages}{1543} (\bibinfo{year}{1998}).
\bibitem{sorensen}
\bibinfo{author}{\bibnamefont{A.~}\bibnamefont{S\o rensen}} {\em et al.},
\bibinfo{journal}{Nature} \textbf{\bibinfo{volume}{409}},
\bibinfo{pages}{63}(\bibinfo{year}{2001}).
\end{thebibliography}

\end{document}